\documentclass[nofootinbib,preprint,preprintnumbers,a4paper,11pt]{revtex4}
\usepackage{epsfig}
\usepackage{footnote}
\usepackage{ulem}
\usepackage{color}
\usepackage{array}
\usepackage{amssymb}
\usepackage{amsmath}
\usepackage{graphicx}
\usepackage{subfigure}
\usepackage{longtable}
\usepackage{verbatim}
\usepackage{amsfonts}
\usepackage{hyperref}
\usepackage{multirow}
\usepackage{cancel}

\begin{document}

\title{From topological amplitude to rescattering dynamics\\ in doubly charmed baryon decays}

\author{Di Wang$^{1}$}\email{wangdi@hunnu.edu.cn}

\address{%
$^1$Department of Physics, Hunan Normal University, and Key Laboratory of Low-Dimensional Quantum Structures and Quantum Control of Ministry of Education, Changsha 410081, China
}

\begin{abstract}

The doubly charmed baryon $\Xi^{++}_{cc}$ was observed by LHCb cooperation in 2017.
The branching fractions of two-body doubly charmed baryon decays were predicted in the framework of rescattering mechanism, and some $SU(3)_F$ relations were investigated in the topological amplitudes.
In this work, we study the correlation between topological diagram at quark level and rescattering triangle diagram at hadron level in the doubly charmed baryon decay.
The completeness of our framework is confirmed from the fact that all the twelve possible structures of meson-baryon scattering appear once each in the the intermediate form between topological diagram and triangle diagram, topological-scattering diagram.
It is found the triangle diagrams derived from the topological diagrams are consistent with the ones derived directly from the chiral Lagrangian.
The relative magnitudes of rescattering contributions in the $C$, $C^\prime$, $E$, $E^\prime$, $P$ and $P^\prime$ diagrams extracted from $SU(3)_F$ symmetry are consistent with the numerical analysis in literature.
Taking $\Xi^{++}_{cc}\to \Xi^+_c\pi^+$, $\Xi^{+}_{cc}\to \Xi^0_c\pi^+$ and $\Xi^{+}_{cc}\to \Xi^+_c\pi^0$ modes as examples, we show the isospin relation is satisfied in terms of triangle diagrams.

\end{abstract}

\maketitle
%{{\tableofcontents}}

\newpage
\section{Introduction}\label{intro}

In 2017, LHCb collaboration observed the doubly charmed baryon $\Xi_{cc}^{++}$ via $\Xi_{cc}^{++}\to \Lambda^+_cK^-\pi^+\pi^+$ decay \cite{LHCb:2017iph}.
Subsequently, the measurement of the lifetime of $\Xi_{cc}^{++}$ and the observation of $\Xi_{cc}^{++}\to \Xi^+_c\pi^+$ were performed \cite{LHCb:2018zpl,LHCb:2018pcs}.
The discovery of $\Xi_{cc}^{++}$ benefits from the theoretical work \cite{Yu:2017zst}, in which the most favorable decay channels of $\Xi_{cc}^{++}$ were pointed out. In Ref.~\cite{Yu:2017zst}, the branching fractions of doubly charmed baryon decays are estimated in the rescattering mechanism, since the QCD-inspired methods do not work well at the scale of charm quark decay.
The rescattering mechanism has been used in heavy mason and baryon hadron decays in literature \cite{Han:2021gkl,Locher:1993cc,Li:1996cj,Dai:1999cs,Li:2002pj,Ablikim:2002ep,Cheng:2004ru,Lu:2005mx,Chen:2002jr}.
And a systematic study on doubly charmed baryon decays in the rescattering mechanism has been performed in \cite{Jiang:2018oak,Han:2021azw}.

In the rescattering mechanism, the doubly charmed baryon first decays into one baryon and one meson via a short-distance emitted amplitude $T^{SD}$. Then the $t$-channel meson-baryon scattering between them serves as the long-distance contributions.
It forms a triangle diagram at hadron level.
There are two different approaches to get the triangle diagrams contributing to one decay channel: calculating the hadron-level Feynman diagrams directly  from the chiral Lagrangian \cite{Jiang:2018oak,Han:2021azw}, or extracting from the topological diagrams \cite{Ablikim:2002ep,Cheng:2004ru}.
In the second method, topological-scattering diagram, the intermediate form between topological diagram and triangle diagram, is used to describe the transition from $T$ diagram to other diagrams such as $E$, $C$ ... etc.
However, the triangle diagrams given by these two methods are not consistent in literature \cite{Han:2021azw,Cheng:2004ru}.
A further study is necessary.

Inspired by the idea of topological diagram expressed in the invariant tensor \cite{He:2018php,He:2018joe,Wang:2020gmn}, we proposed a theoretical framework to associate topological amplitude and rescattering dynamics in heavy meson decays in Ref.~\cite{Wang:2021rhd}.
In this framework, both the triangle diagram and the topological-scattering diagram are expressed in the tensor form.
The coefficients of triangle diagrams can be derived from the quark diagrams.
In this way, the conflict between two approaches to obtain triangle diagrams is solved. The triangle diagrams derived from topological diagrams are the same with the ones derived from the chiral Lagrangian.
In this work, we generalize the theoretical framework proposed in \cite{Wang:2021rhd} to the doubly charmed baryon decays.
It is found the twelve possible structures of meson-baryon scattering appear once each in the topological-scattering diagrams.
And the rescattering contributions in $C$, $C^\prime$, $E$, $E^\prime$, $P$ and $P^\prime$ diagrams have definite proportional relation under the $SU(3)_F$ symmetry.
Taking $\Xi^{++}_{cc}\to \Xi^+_c\pi^+$, $\Xi^{+}_{cc}\to \Xi^0_c\pi^+$ and $\Xi^{+}_{cc}\to \Xi^+_c\pi^0$ modes as examples, we show our framework in detail. One can find the isospin relation between them is held in terms of triangle diagrams.

This paper is organized as follows. In Sec.~\ref{ttor}, we construct the theoretical framework of the relation between topological amplitude and rescattering triangle diagram.
In Sec.~\ref{exam}, the reliability of our method is checked in the
$\Xi^{++}_{cc}\to \Xi^+_c\pi^+$, $\Xi^{+}_{cc}\to \Xi^0_c\pi^+$ and $\Xi^{+}_{cc}\to \Xi^+_c\pi^0$ modes.
And Sec.~\ref{sum} is a short summary.

\section{From topological diagram to triangle diagram}\label{ttor}
\begin{figure}
  \centering
  \includegraphics[width=14cm]{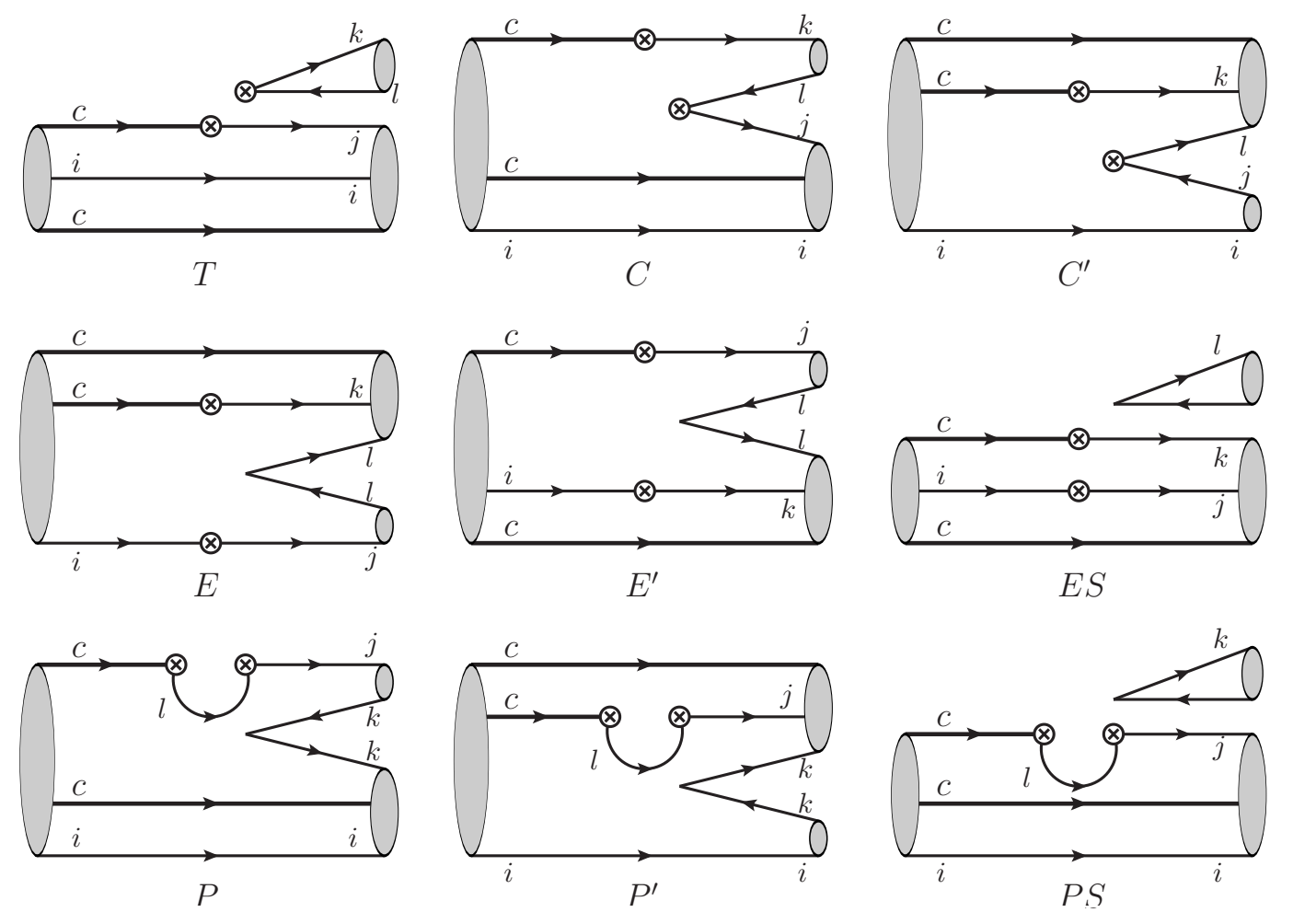}
  \caption{Topological diagrams contributing to the $\mathcal{B}_{cc}\to \mathcal{B}_{c\overline3}M$ decays in the Standard Model.}\label{topo}
\end{figure}
In this section, we write the topological-scattering diagram and triangle diagram in the tensor form and analyze the relation among topological diagram, topological-scattering diagram and triangle diagram.
In the $SU(3)$ picture, pseudoscalar meson nonet $|M^{i}_{j} \rangle$ is expressed as
\begin{eqnarray}\label{a1}
 |M^i_j\rangle =  \left( \begin{array}{ccc}
   \frac{1}{\sqrt 2} |\pi^0\rangle +  \frac{1}{\sqrt 6} |\eta_8\rangle,    & |\pi^+\rangle,  & |K^+\rangle \\
   | \pi^-\rangle, &   - \frac{1}{\sqrt 2} |\pi^0\rangle+ \frac{1}{\sqrt 6} |\eta_8\rangle,   & |K^0\rangle \\
   | K^- \rangle,& |\overline K^0\rangle, & -\sqrt{\frac{2}{3}}|\eta_8\rangle \\
  \end{array}\right) +  \frac{1}{\sqrt 3} \left( \begin{array}{ccc}
   |\eta_1\rangle,    & 0,  & 0 \\
    0, &  |\eta_1\rangle,   & 0 \\
   0, & 0, & |\eta_1\rangle \\
  \end{array}\right),
\end{eqnarray}
where $i$ is row index and $j$ is column index.
The vector meson nonet is
\begin{eqnarray}\label{v}
 |V\rangle ^i_j=  \left( \begin{array}{ccc}
   \frac{1}{\sqrt 2} |\rho^0\rangle+  \frac{1}{\sqrt 2} |\omega\rangle,    & |\rho^+\rangle,  & |K^{*+}\rangle \\
    |\rho^-\rangle, &   - \frac{1}{\sqrt 2} |\rho^0\rangle+ \frac{1}{\sqrt 2} |\omega\rangle,   & |K^{*0}\rangle \\
    |K^{*-}\rangle, & |\overline K^{*0}\rangle, & |\phi\rangle \\
  \end{array}\right).
\end{eqnarray}
The doubly charmed triplet baryon is expressed as
\begin{align}
  |\mathcal{B}_{cc}\rangle  = (|\Xi_{cc}^{++}(ccu)\rangle,\,\,|\Xi_{cc}^{+}(ccd)\rangle,\,\, |\Omega_{cc}^{+}(ccs)\rangle ).
\end{align}
The charmed anti-triplet baryon is expressed as
\begin{eqnarray}
 |\mathcal{B}_{c\overline 3}\rangle=  \left( \begin{array}{ccc}
   0   & |\Lambda_c^+\rangle  & |\Xi_c^+\rangle \\
    -|\Lambda_c^+\rangle &   0   & |\Xi_c^0\rangle \\
    -|\Xi_c^+\rangle & -|\Xi_c^0\rangle & 0 \\
  \end{array}\right).
\end{eqnarray}
The amplitude of doubly charmed baryon decays into a charmed anti-triplet baryon and a light meson in the Standard Model (SM) can be expressed as sum of invariant tensors,
\begin{align}\label{amp}
  \mathcal{A}(\mathcal{B}_{cc}\to \mathcal{B}_{c\overline3}M) = & \,T\,(\mathcal{B}_{cc})_iH^l_{kj}M^k_l(\overline{ \mathcal{B}}_{c\overline3})^{ij} + C\,(\mathcal{B}_{cc})_iH^l_{jk}M^k_l(\overline{ \mathcal{B}}_{c\overline3})^{ij} + C^\prime\,(\mathcal{B}_{cc})_iH^j_{lk}M^i_j(\overline{ \mathcal{B}}_{c\overline3})^{lk}\nonumber\\ &+E\,(\mathcal{B}_{cc})_iH^i_{jk}M^j_l(\overline{ \mathcal{B}}_{c\overline3})^{lk}+  E^\prime\,(\mathcal{B}_{cc})_iH^i_{kj}M^j_l(\overline{ \mathcal{B}}_{c\overline3})^{lk}+  ES\,(\mathcal{B}_{cc})_iH^i_{jk}M^l_l(\overline{ \mathcal{B}}_{c\overline3})^{jk}\nonumber\\
    & + P\,(\mathcal{B}_{cc})_iH^l_{jl}M^j_k(\overline{ \mathcal{B}}_{c\overline3})^{ik}+ P^\prime\,(\mathcal{B}_{cc})_iH^l_{jl}M^i_k(\overline{ \mathcal{B}}_{c\overline3})^{jk}\nonumber\\
    &+PS\,(\mathcal{B}_{cc})_iH^l_{jl}M^k_k(\overline{ \mathcal{B}}_{c\overline3})^{ij}.
\end{align}
If the index-contraction is understood as quark flowing, each term in Eq.~\eqref{amp} is a topological diagram.
The topological diagrams contribute to $\mathcal{B}_{cc}\to \mathcal{B}_{c\overline3}M$ are listed in Fig.~\ref{topo}.
The first five diagrams, $T$, $C$, $C^\prime$, $E$ and $E^\prime$, are tree-level topological diagrams have been analyzed in literatures such as \cite{Jiang:2018oak,Han:2021azw}.
$ES$ is the singlet contribution.
The last three diagrams are quark-loop contributions.
There is a sign arbitrariness in the definition of topological diagram in Eq.~\eqref{amp}.
For example, if we define the $T$ amplitude as $(\mathcal{B}_{cc})_iH^l_{kj}M^k_l(\overline{ \mathcal{B}}_{c\overline3})^{ji}$, an additional minus sign will appear in $T$ amplitude because of $(\overline{\mathcal{B}}_{c\overline3})^{ij}=-(\overline{\mathcal{B}}_{c\overline3})^{ji}$.
\begin{figure}
  \centering
  \includegraphics[width=10cm]{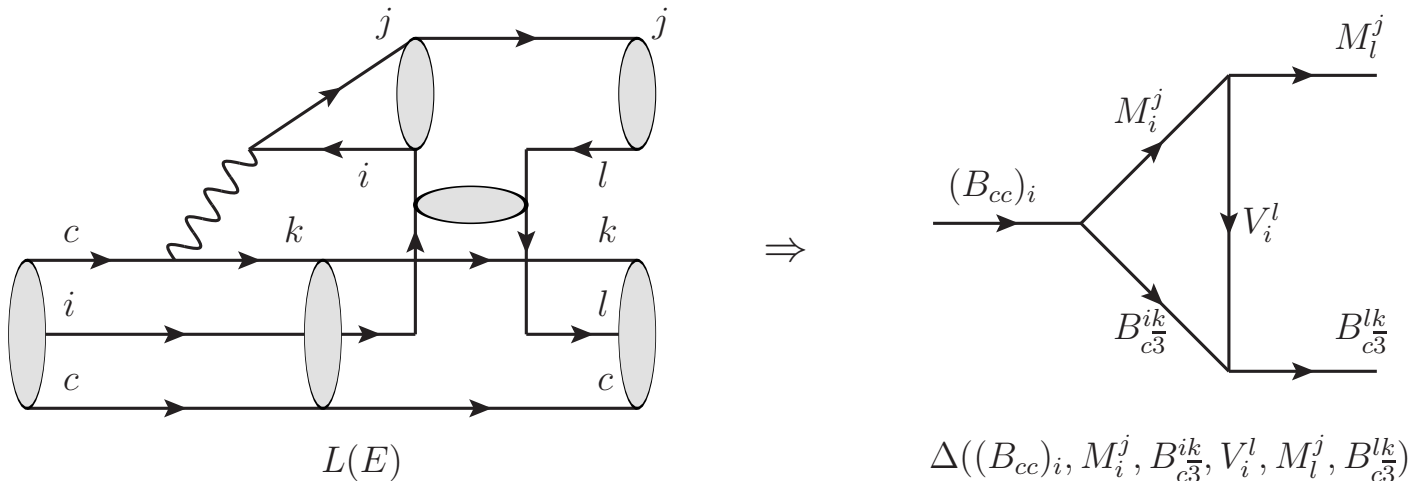}
  \caption{Topological-scattering diagram and triangle diagram in $T\,\Rightarrow\,E$ transition.}\label{E}
\end{figure}

In the factorization approach, amplitude $T$ is dominated by factorizable contribution, $T^{SD}$.
Since $C^{SD}$ is the Fierz transformation of $T^{SD}$, the factorizable part in the $C$ amplitude is also important.
The factorizable contributions $T^{SD}$ and $C^{SD}$ can be parameterized as the decay constant of the emitted mesons and the other is expressed as the transition
form factors\footnote{Please see literature such as Ref.~\cite{Han:2021azw} for details.}.
In the final state interaction (FSI) framework, the non-factorable QCD effects can be modeled as an exchange of one particle between two particles generated from the tree emitted amplitudes, $T^{SD}$ and $C^{SD}$.
There are $s$-channel and $t$-channel contributions in the final state interaction, or referred to as resonance and rescattering contributions respectively.
In this work, we focus on the $t$-channel FSI contribution.
It forms a triangle diagram at hadron level, and can be derived from topological diagram via the topological-scattering diagram.
In the rest of this section, we will study the relation between the topological diagram, topological-scattering diagram and triangle diagram and give physical consequences.
The factorizable contribution of the $C$ diagram is suppressed by the color factor at charm scale with the effective Wilson coefficient $a_2(m_c)=C_1(m_c)+C_2(m_c)/N_c$.
So we neglect the factorizable contribution $C^{SD}$ and only analyze the rescattering contribution arisen from $T^{SD}$ just like Ref.~\cite{Han:2021azw}.

Following Ref.~\cite{Wang:2021rhd}, we express the topological-scattering diagram and triangle diagram in the tensor form, taking $E$ diagram as an example.
The topological-scattering diagram of $T\Rightarrow E$ transition forms a triangle diagram at hadron level, see Fig.~\ref{E}. Here the superscript "SD" in $T$ has been omitted for convenience.
In the tensor form of topological diagram, $T$ diagram is written as $(\mathcal{B}_{cc})_q  H_{mn}^pM^{m}_p  (\overline{ \mathcal{B}}_{c\overline3})^{qn}$, $E$ diagram is written as $(\mathcal{B}_{cc})_i  H_{jk}^iM^{j}_l (\overline{ \mathcal{B}}_{c\overline3})^{lk}$. The $T\Rightarrow E$ transition can be written as
\begin{align}\label{le}
L(E)[i,j,k,l]\,\,=\,\, (\mathcal{B}_{cc})_q  H_{mn}^pM^{m}_p  (\overline{ \mathcal{B}}_{c\overline3})^{qn}\,\cdot\,M^p_mV^l_pM^j_l\,\cdot\,( \mathcal{B}_{c\overline3})_{qn}V^q_l(\overline{ \mathcal{B}}_{c\overline3})^{lk}\,\cdot\,
\delta_{ip}\delta_{jm}\delta_{kn}\,\cdot\,\delta_{iq}.
\end{align}
$L(E)[i,j,k,l]$ is a topological-scattering diagram. It can also be understood as a triangle diagram.
The $T$ diagram in the left is the weak vertex of triangle diagram. The $MVM$ vertex is a meson-meson scattering vertex and $( \mathcal{B}_{c\overline3})V(\overline{ \mathcal{B}}_{c\overline3})$ vertex is a meson-baryon scattering vertex.
The index contractions of $M^p_mM^m_p $,  $V^l_i V^i_l$ and $(\overline{ \mathcal{B}}_{c\overline3})^{qn}(\mathcal{B}_{c\overline3})_{qn}$ are three propagators.
The kronecker symbols are used to set $H_{mn}^p=H_{jk}^i$ and $(\mathcal{B}_{cc})_q=(\mathcal{B}_{cc})_i$.
We only consider the vector meson and charmed anti-triplet baryon exchanges here, i.e., $\mathcal{B}_{cc}\to M\mathcal{B}_{c\overline3}  \to M\mathcal{B}_{c\overline3} $ via exchanging a vector meson or charmed anti-triplet baryon.
For other processes such as $\mathcal{B}_{cc}\to V\mathcal{B}_{c\overline3}\to M\mathcal{B}_{c\overline3}$, $\mathcal{B}_{cc}\to M\mathcal{B}_{c6}\to M\mathcal{B}_{c\overline3}$, ... are similar to the case of $\mathcal{B}_{cc}\to M\mathcal{B}_{c\overline3}  \to M\mathcal{B}_{c\overline3} $.
We will not present details in this work.
Besides, please notice the order of indies of meson-baryon vertex. Due to the  antisymmetric light quarks in the charmed anti-triplet baryon, we have
\begin{align}\label{ver}
(B_{c\overline 3})_{ij}V^i_k(\overline B_{c\overline 3})^{kl}=-(B_{c\overline 3})_{ij}V^i_l(\overline B_{c\overline 3})^{kl}=(B_{c\overline 3})_{ij}V^j_l(\overline B_{c\overline 3})^{kl}=-(B_{c\overline 3})_{ij}V^j_k(\overline B_{c\overline 3})^{kl}.
\end{align}
It guarantees the sign arbitrariness of topological diagram cannot affect the sign of triangle diagram.

\begin{figure}
  \centering
  \includegraphics[width=10cm]{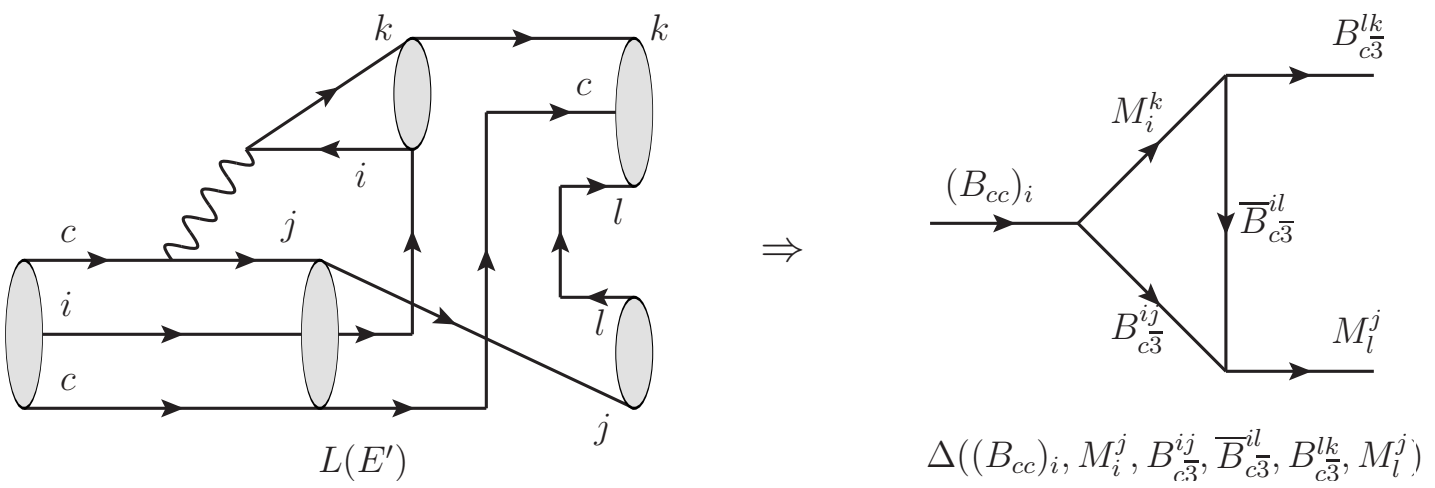}
  \caption{Topological-scattering diagram and triangle diagram in $T\,\Rightarrow\,E^\prime$ transition.}\label{E2}
\end{figure}
\begin{figure}
  \centering
  \includegraphics[width=10cm]{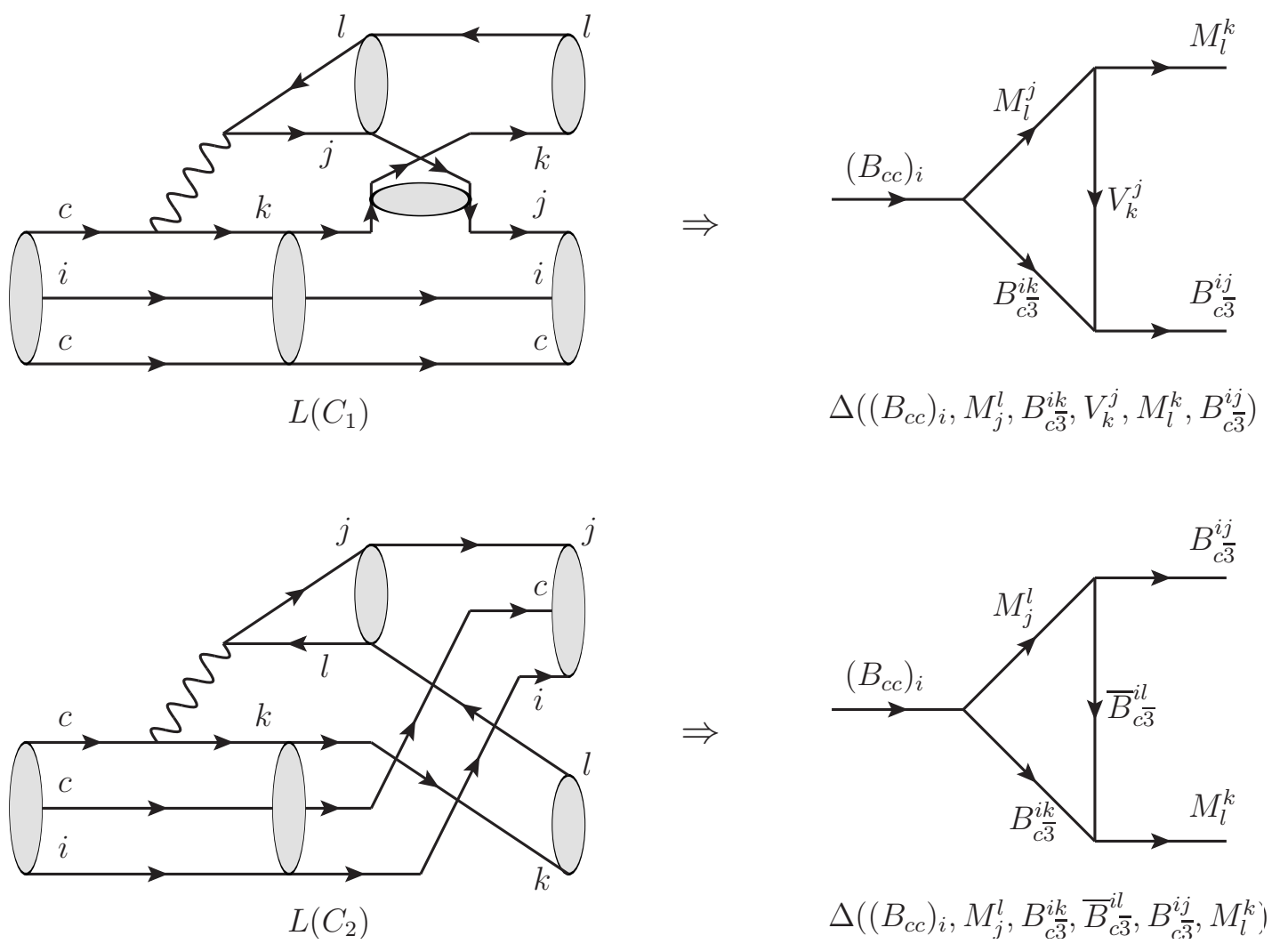}
  \caption{Topological-scattering diagram and triangle diagram in $T\,\Rightarrow\,C$ transition.}\label{C}
\end{figure}
\begin{figure}
  \centering
  \includegraphics[width=10cm]{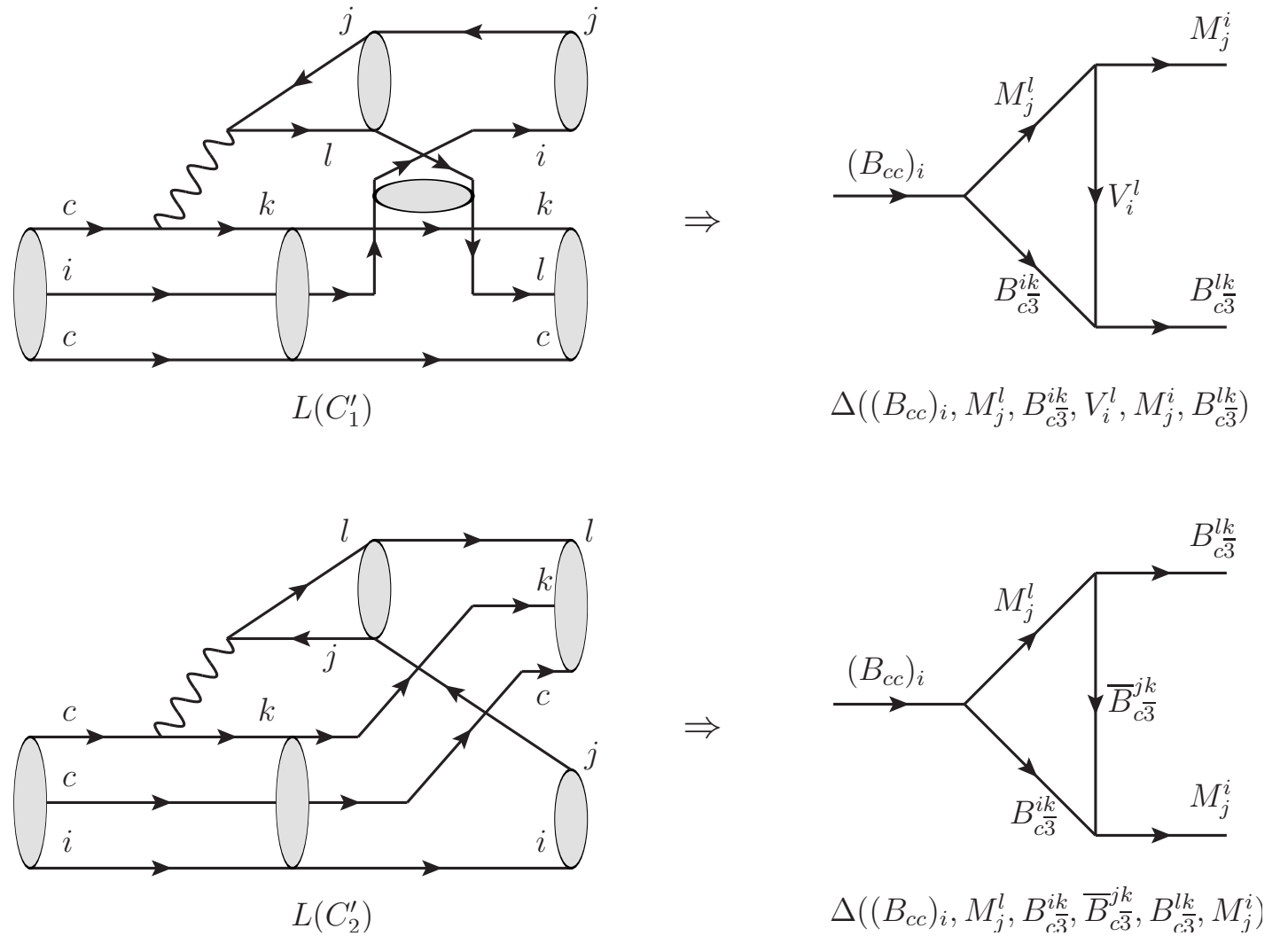}
  \caption{Topological-scattering diagram and triangle diagram in $T\,\Rightarrow\,C^\prime$ transition.}\label{C2}
\end{figure}
\begin{figure}
  \centering
  \includegraphics[width=10cm]{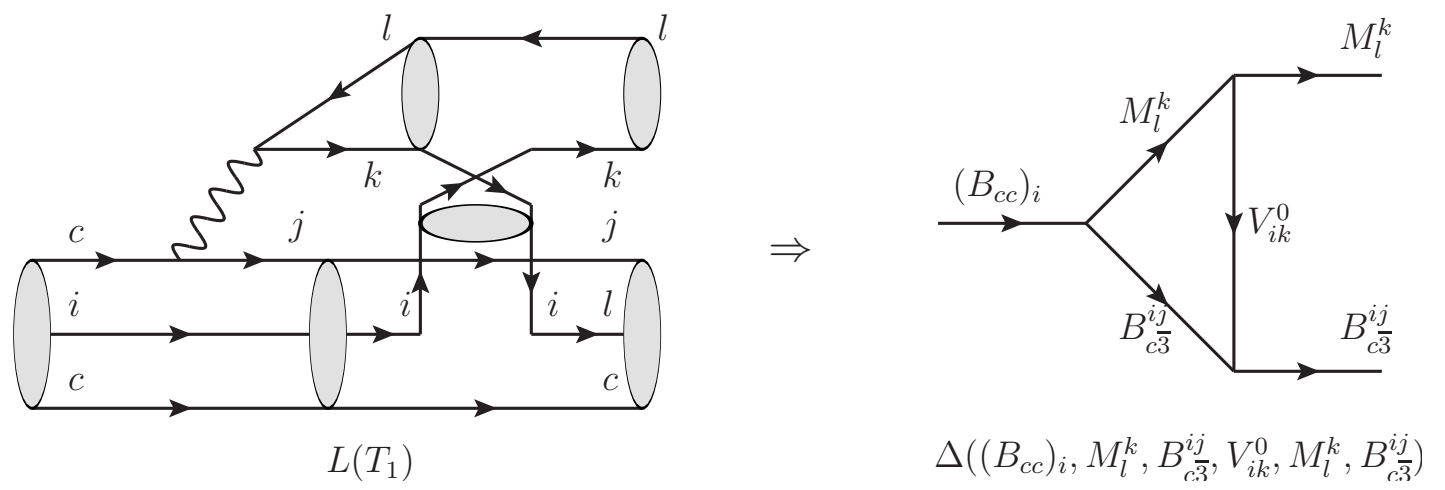}\\\vspace{0.5cm}
  \includegraphics[width=10cm]{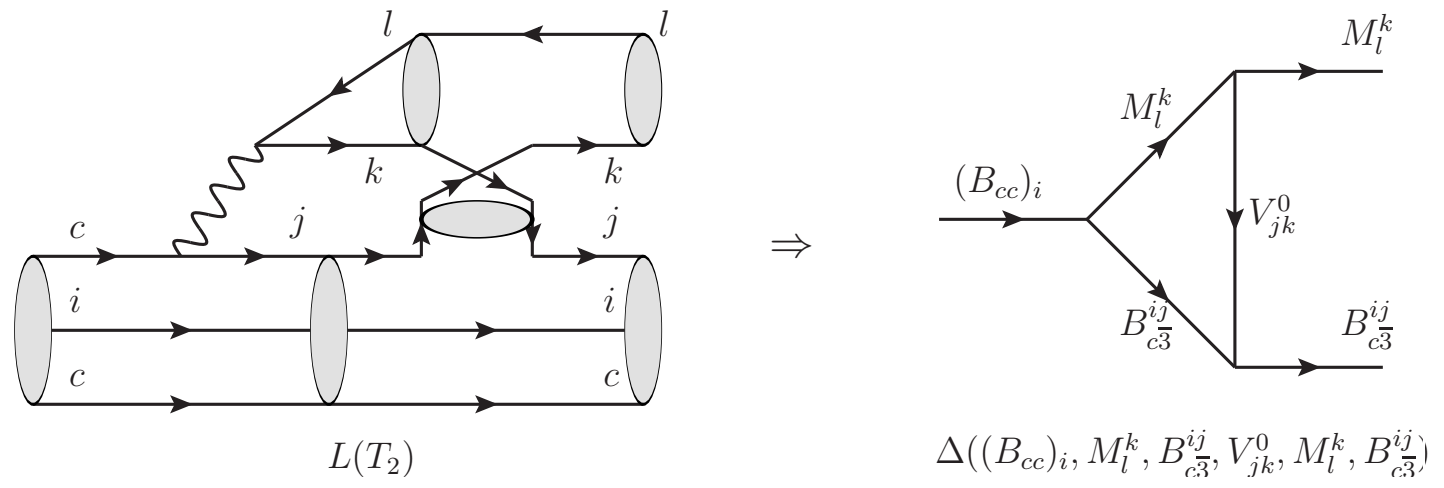}\\\vspace{0.5cm}
  \includegraphics[width=10cm]{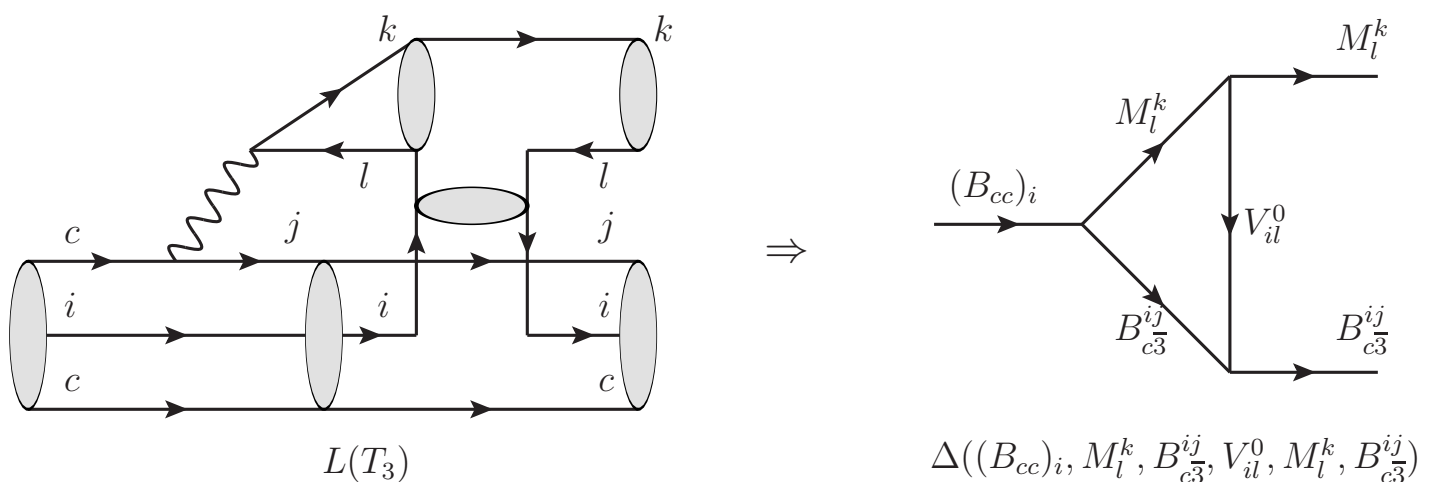}\\\vspace{0.5cm}
  \includegraphics[width=10cm]{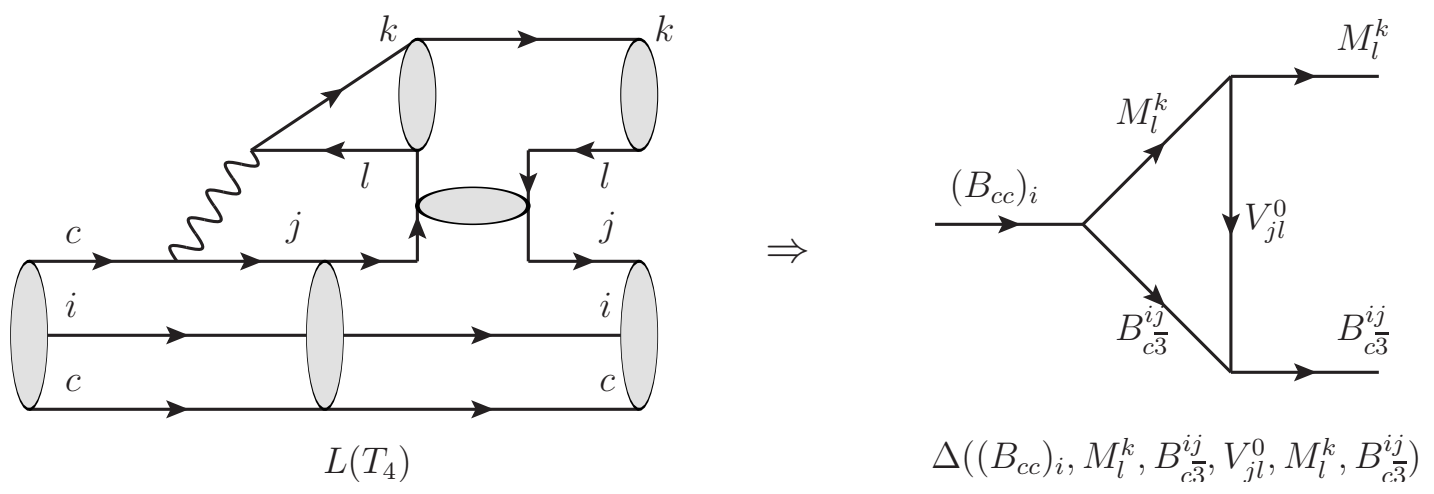}
  \caption{Topological-scattering diagram and triangle diagram in $T\,\Rightarrow\,T$ transition.}\label{T}
\end{figure}
\begin{figure}
  \centering
  \includegraphics[width=10cm]{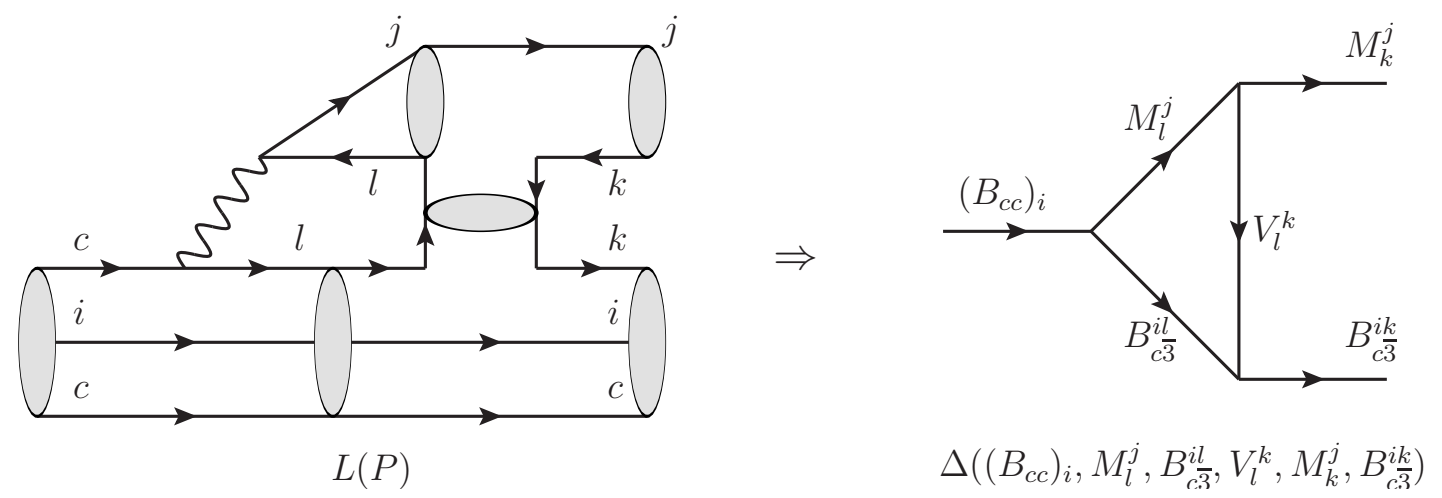}
  \caption{Topological-scattering diagram and triangle diagram in $T\,\Rightarrow\,P$ transition.}\label{P}
\end{figure}
\begin{figure}
  \centering
  \includegraphics[width=10cm]{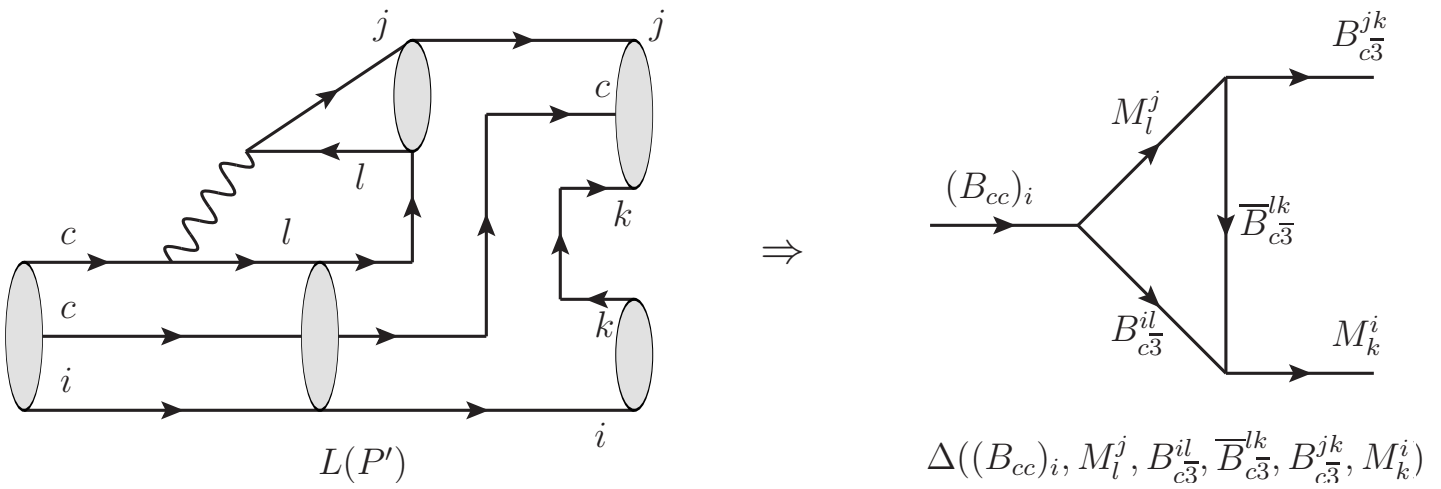}
  \caption{Topological-scattering diagram and triangle diagram in $T\,\Rightarrow\,P^\prime$ transition.}\label{P2}
\end{figure}
For the completeness of our theoretical framework, we list all the tensor structures in $T\Rightarrow E^\prime,\,C,\,C^\prime,\,T,\,P,\,P^\prime$ transitions. \\
$T\Rightarrow E^\prime$:
\begin{align}
L(E^\prime)[i,j,k,l]\,\,=\,\, &(\mathcal{B}_{cc})_q  H_{mn}^pM^{m}_p  (\overline{ \mathcal{B}}_{c\overline3})^{qn}\,\cdot\,M^p_m (\mathcal{B}_{c\overline3})_{pl} (\overline{ \mathcal{B}}_{c\overline3})^{lk}\,\cdot\,(\mathcal{B}_{c\overline3})_{qn}(\overline{ \mathcal{B}}_{c\overline3})^{ql}M^j_l\nonumber\\&\,\cdot\,
\delta_{ip}\delta_{km}\delta_{jn}\,\cdot\,\delta_{iq}.
\end{align}
$T\Rightarrow C$:
\begin{align}
L(C)[i,j,k,l]_1\,\ =\,\, &(\mathcal{B}_{cc})_q  H_{mn}^pM^{m}_p  (\overline{ \mathcal{B}}_{c\overline3})^{qn}\,\cdot\,M^p_mV^m_kM^k_l\,\cdot\,( \mathcal{B}_{c\overline3})_{qn}V^n_j(\overline{ \mathcal{B}}_{c\overline3})^{ij}\,\cdot\,
\delta_{lp}\delta_{jm}\delta_{kn}\,\cdot\,\delta_{iq},\\
L(C)[i,j,k,l]_2\,\, =\,\, &(\mathcal{B}_{cc})_q  H_{mn}^pM^{m}_p  (\overline{ \mathcal{B}}_{c\overline3})^{qn}\,\cdot\,M^p_m (\mathcal{B}_{c\overline3})_{ip} (\overline{ \mathcal{B}}_{c\overline3})^{ij}\,\cdot\,(\mathcal{B}_{c\overline3})_{qn}(\overline{ \mathcal{B}}_{c\overline3})^{ql}M^k_l\nonumber\\&\,\cdot\,
\delta_{lp}\delta_{jm}\delta_{kn}\,\cdot\,\delta_{iq}.
\end{align}
$T\Rightarrow C^\prime$:
\begin{align}
L(C^\prime)[i,j,k,l]_1\,\ =\,\, &(\mathcal{B}_{cc})_q  H_{mn}^pM^{m}_p  (\overline{ \mathcal{B}}_{c\overline3})^{qn}\,\cdot\,M^p_mV^m_iM^i_j\,\cdot\,( \mathcal{B}_{c\overline3})_{qn}V^q_l(\overline{ \mathcal{B}}_{c\overline3})^{lk}\,\cdot\,
\delta_{jp}\delta_{lm}\delta_{kn}\,\cdot\,\delta_{iq},\\
L(C^\prime)[i,j,k,l]_2\,\, =\,\, &(\mathcal{B}_{cc})_q  H_{mn}^pM^{m}_p  (\overline{ \mathcal{B}}_{c\overline3})^{qn}\,\cdot\,M^p_m (\mathcal{B}_{c\overline3})_{pk} (\overline{ \mathcal{B}}_{c\overline3})^{lk}\,\cdot\,(\mathcal{B}_{c\overline3})_{qn}(\overline{ \mathcal{B}}_{c\overline3})^{jn}M^i_j\nonumber\\&\,\cdot\,
\delta_{jp}\delta_{lm}\delta_{kn}\,\cdot\,\delta_{iq}.
\end{align}
$T\Rightarrow T$:
\begin{align}
L(T)[i,j,k,l]_1\,\,=\,\, & (\mathcal{B}_{cc})_q  H_{mn}^pM^{m}_p  (\overline{ \mathcal{B}}_{c\overline3})^{qn}\,\cdot\,M^p_mV^m_kM^k_l\,\cdot\,(\mathcal{B}_{c\overline3})_{qn}V^q_i(\overline{ \mathcal{B}}_{c\overline3})^{ij}\,\cdot\,
\delta_{lp}\delta_{km}\delta_{jn}\,\cdot\,\delta_{iq},\\
L(T)[i,j,k,l]_2\,\,=\,\, & (\mathcal{B}_{cc})_q  H_{mn}^pM^{m}_p  (\overline{ \mathcal{B}}_{c\overline3})^{qn}\,\cdot\,M^p_mV^m_kM^k_l\,\cdot\,(\mathcal{B}_{c\overline3})_{qn}V^n_j(\overline{ \mathcal{B}}_{c\overline3})^{ij}\,\cdot\,
\delta_{lp}\delta_{km}\delta_{jn}\,\cdot\,\delta_{iq},\\
L(T)[i,j,k,l]_3\,\,=\,\, & (\mathcal{B}_{cc})_q  H_{mn}^pM^{m}_p  (\overline{ \mathcal{B}}_{c\overline3})^{qn}\,\cdot\,M^p_mV^l_pM^k_l\,\cdot\,(\mathcal{B}_{c\overline3})_{qn}V^q_i(\overline{ \mathcal{B}}_{c\overline3})^{ij}\,\cdot\,
\delta_{lp}\delta_{km}\delta_{jn}\,\cdot\,\delta_{iq},\\
L(T)[i,j,k,l]_4\,\,=\,\, & (\mathcal{B}_{cc})_q  H_{mn}^pM^{m}_p  (\overline{ \mathcal{B}}_{c\overline3})^{qn}\,\cdot\,M^p_mV^l_pM^k_l\,\cdot\,(\mathcal{B}_{c\overline3})_{qn}V^n_j(\overline{ \mathcal{B}}_{c\overline3})^{ij}\,\cdot\,
\delta_{lp}\delta_{km}\delta_{jn}\,\cdot\,\delta_{iq}.
\end{align}
$T\Rightarrow P$:
\begin{align}
L(P)[i,j,k,l]\,\,=\,\, (\mathcal{B}_{cc})_q  H_{mn}^pM^{m}_p  (\overline{ \mathcal{B}}_{c\overline3})^{qn}\,\cdot\,M^p_mV^k_pM^j_k\,\cdot\,( \mathcal{B}_{c\overline3})_{qn}V^n_k(\overline{ \mathcal{B}}_{c\overline3})^{ik}\,\cdot\,
\delta_{lp}\delta_{jm}\delta_{ln}\,\cdot\,\delta_{iq}.
\end{align}
$T\Rightarrow P^\prime$:
\begin{align}
L(P^\prime)[i,j,k,l]\,\,=\,\, &(\mathcal{B}_{cc})_q  H_{mn}^pM^{m}_p  (\overline{ \mathcal{B}}_{c\overline3})^{qn}\,\cdot\,M^p_m (\mathcal{B}_{c\overline3})_{pk} (\overline{ \mathcal{B}}_{c\overline3})^{jk}\,\cdot\,(\mathcal{B}_{c\overline3})_{qn}(\overline{ \mathcal{B}}_{c\overline3})^{nk}M^i_k\nonumber\\&\,\cdot\,
\delta_{lp}\delta_{jm}\delta_{ln}\,\cdot\,\delta_{iq}.
\end{align}
The triangle diagrams constructed by the topological-scattering diagrams in $T\Rightarrow E^\prime$, $T\Rightarrow C$, $T\Rightarrow C^\prime$, $T\Rightarrow T$, $T\Rightarrow P$ and $T\Rightarrow P^\prime$ transitions are shown in Figs.~\ref{E2} $\sim$ \ref{P2}, respectively.
As pointed out in Ref.~\cite{Wang:2021rhd}, the non-perturbative effects in the $ES$ and $PS$ diagrams cannot be modeled into triangle diagram because they can be divided into two unconnected parts by cutting off gluon propagators.

\begin{figure}
  \centering
  \includegraphics[width=10cm]{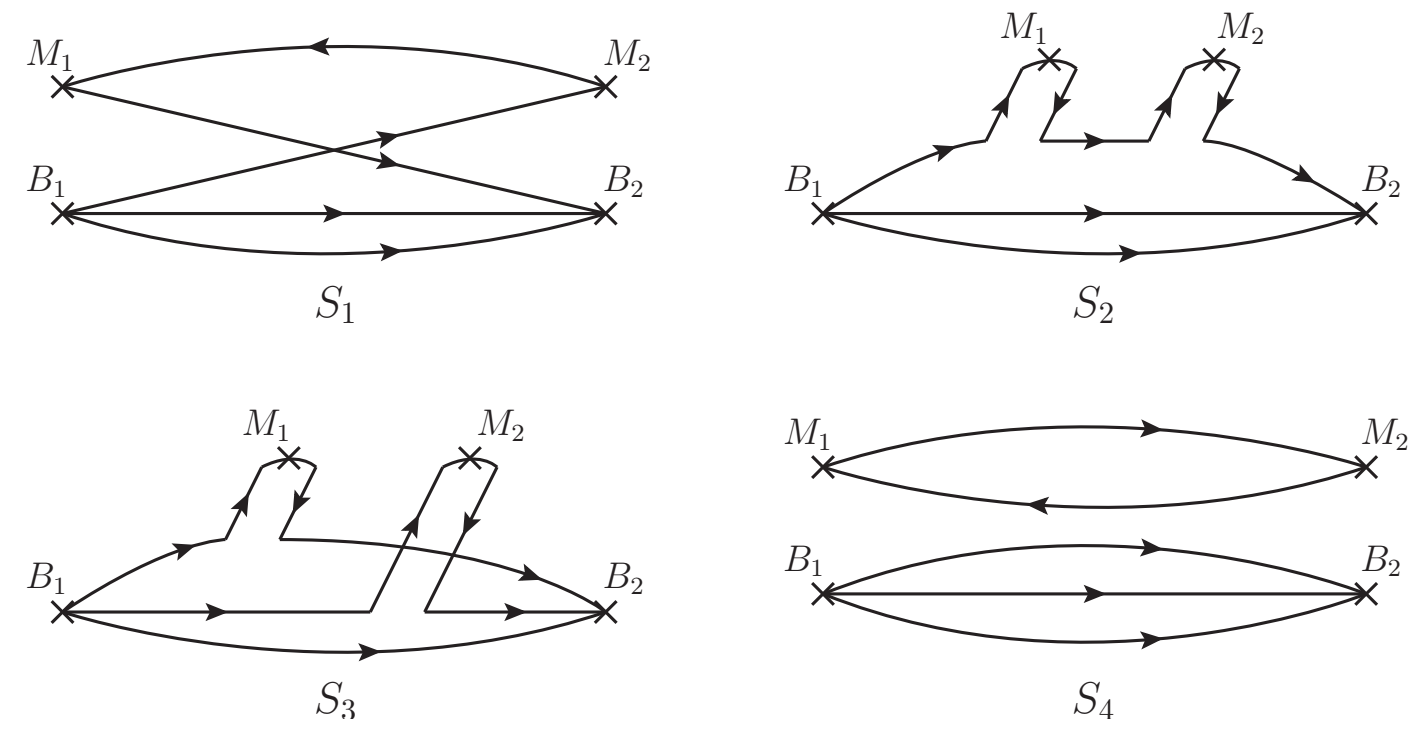}
  \caption{Sketches of meson-baryon scattering in the long-distance contributions of doubly charmed baryon decays into a charmed anti-triplet baryon and a light meson.}\label{scatter}
\end{figure}
There are twelve different sub-structures of meson-baryon scattering in Figs.~\ref{E} $\sim$ \ref{P2}. All of them can be summarized as Fig.~\ref{scatter}.
$S_1$ in Fig.~\ref{scatter} represents the meson-baryon scattering induced by a quark exchange between meson and baryon.
There are four different choices of exchanging a light quark between a light meson and a heavy baryon, corresponding to the meson-baryon scattering in topological-scattering diagrams $L(C)_1$, $L(C^\prime)_1$, $L(E)$ and $L(P)$, respectively.
$S_2$ in Fig.~\ref{scatter} represents the meson-baryon scattering induced by inserting two meson currents in one valence quark of baryon. There are two different choices in light meson-heavy baryon scattering, corresponding to the scattering in topological-scattering diagrams $L(C)_2$ and $L(C^\prime)_2$.
$S_3$ in Fig.~\ref{scatter} represents the meson-baryon scattering induced by inserting two meson currents in two valence quarks of baryon.
There are two different choices in light meson-heavy baryon scattering, corresponding to the scattering in topological-scattering diagrams $L(E^\prime)$ and $L(P^\prime)$.
$S_4$ in Fig.~\ref{scatter} represents the meson-baryon scattering induced by exchanging a neutral meson propagator without quark exchange or meson current inserting.
There are four different choices in light meson-heavy baryon scattering, corresponding to the scattering in topological-scattering diagrams $L(T)_1$, $L(T)_2$, $L(T)_3$, $L(T)_4$, respectively.
In the end, the topological-scattering diagrams in $T \Rightarrow C^{(\prime)}$, $T \Rightarrow E^{(\prime)}$, $T \Rightarrow P^{(\prime)}$ and $T \Rightarrow T$ transitions cover all the twelve possible structures of meson-baryon scattering without repetition.
Thereby, the topological-scattering diagrams listed in Figs.~\ref{E} $\sim$ \ref{P2} are complete.

\section{Examples and discussions}\label{exam}
In this section, we take the $\Xi^{++}_{cc}\to \Xi^+_c\pi^+$, $\Xi^{+}_{cc}\to \Xi^0_c\pi^+$ and $\Xi^{+}_{cc}\to \Xi^+_c\pi^0$ modes as examples to illustrate the reliability of the method proposed in last section.
The topological amplitude of $\Xi^{++}_{cc}\to \Xi^+_c\pi^+$ decay is $T+C^\prime$.
The rescattering contributions modeled by triangle diagram at hadron level can be written as
\begin{align}\label{x4}
L(C^\prime)_1[u,d,s,u]\,\,&=\,\, -\frac{1}{2}\Delta(\Xi^{++}_{cc},\pi^+,\Xi^+_c,\rho^0,\pi^+,\Xi^+_c)
-\frac{1}{2}\Delta(\Xi^{++}_{cc},\pi^+,\Xi^+_c,\omega,\pi^+,\Xi^+_c),\\
L(C^\prime)_2[u,d,s,u]\,\,&=\,\, \Delta(\Xi^{++}_{cc},\pi^+,\Xi^+_c,\overline \Xi^0_c,\Xi^+_c,\pi^+),\\
\label{xx1}
L(T)_3[u,s,u,d]\,\,&=\,\, -\frac{1}{2}\Delta(\Xi^{++}_{cc},\pi^+,\Xi^+_c,\rho^0,\pi^+,\Xi^+_c)
+\frac{1}{2}\Delta(\Xi^{++}_{cc},\pi^+,\Xi^+_c,\omega,\pi^+,\Xi^+_c).
\end{align}
In the topological-scattering diagram $L(C^\prime)_1[u,d,s,u]$, the quark constituent of vector propagator is $u\overline u$. According to Eq.~\eqref{v}, $u\overline u =  |\rho^0\rangle/\sqrt 2+  |\omega\rangle/\sqrt 2$.
There are two strong vertexes in the triangle diagram.
The coefficient $1/2$ is induced from multiplying $1/\sqrt 2$ two times, $(1/\sqrt 2) \times (1/\sqrt 2) = 1/2$.
The minus sign before $1/2$ arises from the cross in $L(C^\prime)_1[u,d,s,u]$ because of the commutator in the effective chiral lagrangian of meson-meson scattering \cite{Wang:2021rhd}.
In $T\Rightarrow T$ transition, the topological-scattering diagram $L(T)_1[u,s,u,d]$ vanishes due to the Pauli exclusion principle \cite{Wang:2021rhd} and hence only $L(T)_3[u,s,u,d]$ is left.
In topological-scattering diagram $L(T)_3[u,s,u,d]$, the quark constituent of vector propagator is $u\overline u/d\overline d$. $u\overline u =  |\rho^0\rangle/\sqrt 2+  |\omega\rangle/\sqrt 2$, $d\overline d =  -|\rho^0\rangle/\sqrt 2+  |\omega\rangle/\sqrt 2$ and hence the propagator $V^0_{il} = -\rho^0/2+\omega/2$.
Summing the $T$ and $C^\prime$ amplitudes, the rescattering contributions in the $\Xi^{++}_{cc}\to \Xi^+_c\pi^+$ decay is
\begin{align}\label{x1}
\mathcal{A}_L(\Xi^{++}_{cc}\to \Xi^+_c\pi^+)&= L(C^\prime)_1[u,d,s,u]+L(C^\prime)_2[u,d,s,u]+L(T)_3[u,s,u,d]\nonumber\\& = -\Delta(\Xi^{++}_{cc},\pi^+,\Xi^+_c,\rho^0,\pi^+,\Xi^+_c)+\Delta(\Xi^{++}_{cc},\pi^+,\Xi^+_c,\overline \Xi^0_c,\Xi^+_c,\pi^+).
\end{align}
Notice that the contributions associated with $\omega\pi\pi$ vertex cancel each other.

Similarly, the topological amplitude of $\Xi^{+}_{cc}\to \Xi^+_c\pi^0$ decay is $\frac{1}{\sqrt{2}}(E-C^\prime)$.
The rescattering contributions include
\begin{align}
-\frac{1}{\sqrt{2}}L(C^\prime)_1[d,d,s,u]\,\,&=\,\, \frac{1}{\sqrt{2}}\Delta(\Xi^{+}_{cc},\pi^+,\Xi^0_c,\rho^+,\pi^0,\Xi^+_c),\\
-\frac{1}{\sqrt{2}}L(C^\prime)_2[d,d,s,u]\,\,&=\,\, -\frac{1}{\sqrt{2}}\Delta(\Xi^{+}_{cc},\pi^+,\Xi^0_c,\overline \Xi^0_c,\Xi^+_c,\pi^0),\\
\frac{1}{\sqrt{2}}L(E)[d,u,s,u]\,\,&=\,\, \frac{1}{\sqrt{2}}\Delta(\Xi^{+}_{cc},\pi^+,\Xi^0_c,\rho^+,\pi^0,\Xi^+_c).
\end{align}
Summing the $C^\prime$ and $E$ amplitudes, we have
\begin{align}\label{x2}
\mathcal{A}_L(\Xi^{+}_{cc}\to \Xi^+_c\pi^0)=&-\frac{1}{\sqrt{2}}(L(C^\prime)_1[d,d,s,u]+L(C^\prime)_2[d,d,s,u]-L(E)[d,u,s,u])\nonumber\\
=& \sqrt{2}\Delta(\Xi^{+}_{cc},\pi^+,\Xi^0_c,\rho^+,\pi^0,\Xi^+_c)-\frac{1}{\sqrt{2}}\Delta(\Xi^{+}_{cc},\pi^+,\Xi^0_c,\overline \Xi^0_c,\Xi^+_c,\pi^0).
\end{align}
The topological amplitude of $\Xi^{+}_{cc}\to \Xi^0_c\pi^+$ decay is $T+E$.
The rescattering contributions include
\begin{align}\label{x5}
L(E)[d,u,s,d]\,\,&=\,\, \frac{1}{2}\Delta(\Xi^{+}_{cc},\pi^+,\Xi^0_c,\rho^0,\pi^+,\Xi^0_c)+\frac{1}{2}\Delta(\Xi^{+}_{cc},\pi^+,\Xi^0_c,\omega,\pi^+,\Xi^0_c),\\
L(T)_1[d,s,u,d]\,\,&=\,\, \frac{1}{2}\Delta(\Xi^{+}_{cc},\pi^+,\Xi^0_c,\rho^0,\pi^+,\Xi^0_c)-\frac{1}{2}\Delta(\Xi^{+}_{cc},\pi^+,\Xi^0_c,\omega,\pi^+,\Xi^0_c).
\end{align}
Summing the $T$ and $E$ amplitudes, we have
\begin{align}\label{x3}
\mathcal{A}_L(\Xi^{+}_{cc}\to \Xi^0_c\pi^+)=&L(E)[d,u,s,d]+L(T)_1[d,s,u,d]
= \Delta(\Xi^{+}_{cc},\pi^+,\Xi^0_c,\rho^0,\pi^+,\Xi^0_c).
\end{align}
Again, all the contributions associated with $\omega\pi\pi$ vertex cancel each other.

Under the isospin symmetry, the particles in an isospin multiplet are the same.
Then we have
\begin{align}
\Delta_1&= \Delta(\Xi^{++}_{cc},\pi^+,\Xi^+_c,\rho^0,\pi^+,\Xi^+_c) = \Delta(\Xi^{+}_{cc},\pi^+,\Xi^0_c,\rho^+,\pi^0,\Xi^+_c) = \Delta(\Xi^{+}_{cc},\pi^+,\Xi^0_c,\rho^0,\pi^+,\Xi^0_c), \nonumber\\
  \Delta_2&=\Delta(\Xi^{++}_{cc},\pi^+,\Xi^+_c,\overline \Xi^0_c,\Xi^+_c,\pi^+) = \Delta(\Xi^{+}_{cc},\pi^+,\Xi^0_c,\overline \Xi^0_c,\Xi^+_c,\pi^0).
\end{align}
The decay amplitudes of $\Xi^{++}_{cc}\to \Xi^+_c\pi^+$, $\Xi^{+}_{cc}\to \Xi^+_c\pi^0$ and $\Xi^{+}_{cc}\to \Xi^0_c\pi^+$ channels can be written as
\begin{align}
\mathcal{A}_L(\Xi^{++}_{cc}\to \Xi^+_c\pi^+)& =-\Delta_1+\Delta_2,\qquad \mathcal{A}_L(\Xi^{+}_{cc}\to \Xi^+_c\pi^0)=\sqrt{2}\Delta_1-\frac{1}{\sqrt{2}}\Delta_2, \nonumber\\
\mathcal{A}_L(\Xi^{+}_{cc}\to \Xi^0_c\pi^+)&=\Delta_1.
\end{align}
One can check the isospin relation
\begin{align}
\mathcal{A}(\Xi^{++}_{cc}\to \Xi^+_c\pi^+)+\sqrt{2}\mathcal{A}(\Xi^{+}_{cc}\to \Xi^+_c\pi^0)-\mathcal{A}(\Xi^{+}_{cc}\to \Xi^0_c\pi^+)=0
\end{align}
is satisfied in terms of triangle diagrams.

Because of the cancellation of the neutral propagators such as in Eq.~\eqref{xx1}, the rescattering contributions in the $T$ diagram are zero under the flavor $SU(3)$ symmetry \cite{Wang:2021rhd}, $L(T)_1  = L(T)_2 = L(T)_3 = L(T)_4 =0$.
Considering that there are two different configurations of triangle diagrams, the intermediate exchange particle served by meson or baryon, and all triangle diagrams are the same with same configuration under the flavor $SU(3)$ symmetry, we get two proportional relations between the topological-scattering diagrams:
\begin{align}\label{eq}
L(C)_1= L(C^\prime)_1 = -L(E) = -L(P),\qquad
 L(C)_2&= L(C^\prime)_2 = -L(E^\prime) = -L(P^\prime).
\end{align}
Notice the sign arbitrariness of topological diagram does not affect the sign of triangle diagram because of Eq.~\eqref{ver}. There is a sign arbitrariness of the equations in Eq.~\eqref{eq}.

\begin{figure}
  \centering
  \includegraphics[width=5cm]{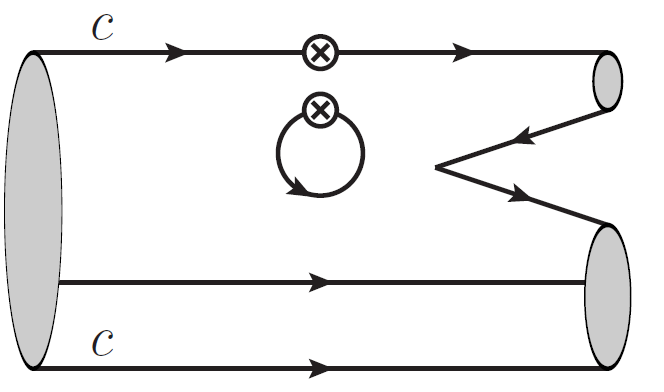}\qquad\qquad
    \includegraphics[width=5cm]{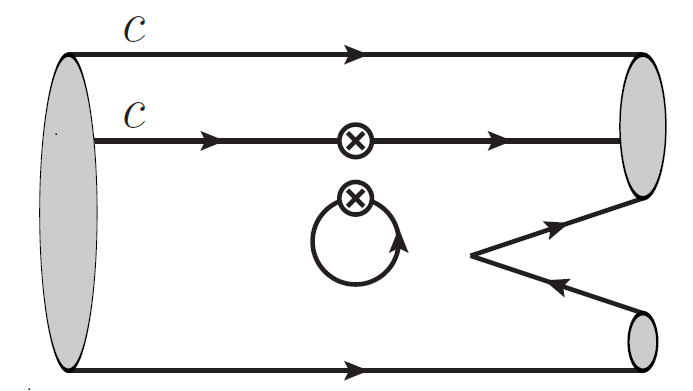}
  \caption{$P_F$ (left) and $P^\prime_F$ (right) diagrams, the Fierz transformations of $P$ and $P^\prime$ diagrams, respectively.}\label{px}
\end{figure}
The rescattering contributions can also be constructed by the two particles emitted from the short-distance $C$ amplitude. In the topological-scattering diagram, it is equivalent to use $C^{SD}$ to replace $T^{SD}$.
The topological-scattering diagram of $C \Rightarrow X$ transition $L^{\prime}(X)$ can be obtained from the topological-scattering diagram of $T \Rightarrow Y$ transition, where $X$, $Y$ are two same or different topological diagrams.
Specifically, the relations between the topological-scattering diagrams arisen from $C^{SD}$ and $T^{SD}$ are summarized to be
\begin{align}\label{csd}
& L^{\prime}(C)_i = \frac{C^{SD}}{T^{SD}}L(T)_i, \qquad L^{\prime}(E^\prime) = \frac{C^{SD}}{T^{SD}}L(E),\qquad L^{\prime}(E) = \frac{C^{SD}}{T^{SD}}L(E^\prime), \qquad L^{\prime}(T)_i = \frac{C^{SD}}{T^{SD}}L(C)_i, \nonumber\\
& L^{\prime}(C^\prime)_i = \frac{C^{SD}}{T^{SD}}L(C^\prime)_i, \qquad L^{\prime}(P) = \frac{C^{SD}}{T^{SD}}L(P_F),\qquad L^{\prime}(P^\prime) = \frac{C^{SD}}{T^{SD}}L(P_F^\prime),
\end{align}
in which $P_F^{(\prime)}$ diagram is the Fierz transformation of $P^{(\prime)}$ diagram, see Fig.~\ref{px}.
In the SM, $P_F^{(\prime)}$ diagram is zero if the tree operators $O_1$ or $Q_2$ is inserted into its weak vertex.
Since $C^{SD}/T^{SD}$ is expected to be suppressed at least by one order, the rescattering contributions arisen from $C^{SD}$ will not affect Eq.~\eqref{eq} heavily.
If the $s$-channel one particle exchange is considered, Eq.~\eqref{eq} will be broken.
But Eq.~\eqref{eq} is still significant as a rough estimation of the long-distance contributions.

In Ref.~\cite{Han:2021azw}, the authors extracted the ratios of topological amplitudes based on the numerical estimation of triangle diagrams in the rescattering mechanism and concluded that $|C^\prime|/|C|\sim |E|/|C|\sim |E^\prime|/|C|\sim \mathcal{O}(1)$.
The same conclusion is also obtained in the soft-collinear effective theory \cite{Mantry:2003uz,Leibovich:2003tw}.
In this work, this conclusion is verified without numerical analysis.
In the large $N_c$ expansion, the meson-baryon scattering $S_1$, $S_2$ and $S_3$ in Fig.~\ref{scatter} are at the same order, $N_c^0$ \cite{Donoghue:1992dd}.
Thereby, all the triangle diagrams in rescattering mechanism have the same order according to Eq.~\eqref{eq}.
The long-distance contributions in topologies $C$, $C^\prime$, $E$, $E^\prime$ are expressed as one or two triangle diagrams and hence comparable.
Except for $C$, $C^\prime$, $E$, $E^\prime$ diagrams, Eq.~\eqref{eq} also indicates that $|P|\sim |P^\prime|\sim |C|$.
Then the relation can be extended to  $|C^\prime|/|C|\sim |E|/|C|\sim |E^\prime|/|C|\sim |P|/|C|\sim |P^\prime|/|C|\sim\mathcal{O}(1)$.
The large $P$ and $P^\prime$ amplitudes could result in a CP violation of the order of $10^{-3}$ in the charmed baryon decays.

\section{Summary}\label{sum}

In this work, we investigated the correlation between topological diagrams at quark level and rescattering dynamics at hadron level in the doubly charmed baryon decays.
It is found the rescattering triangle diagrams derived from topological diagrams are consistent with the ones derived from the chiral Lagrangian.
All the twelve possible structures of meson-baryon scattering appear once each in the topological-scattering diagrams.
The rescattering contributions in $C$, $C^\prime$, $E$, $E^\prime$, $P$ and $P^\prime$ diagrams have definite proportional relation under the $SU(3)_F$ symmetry.
Our framework is checked in the $\Xi^{++}_{cc}\to \Xi^+_c\pi^+$, $\Xi^{+}_{cc}\to \Xi^0_c\pi^+$ and $\Xi^{+}_{cc}\to \Xi^+_c\pi^0$ modes. The isospin relation is satisfied in terms of the triangle diagrams.

\begin{acknowledgements}

This work was supported in part by the National Natural Science Foundation of China
under Grants No.12105099.
\end{acknowledgements}

\end{document}